\documentstyle[preprint,aps]{revtex}

\def\pbar{\rlap\slash p}

\def\overlay#1#2{\setbox0=\hbox{#1}\setbox1=\hbox to \wd0{\hss
#2\hss}#1\hskip -2\wd0\copy1}
\def\lsim{\mathrel{\rlap{\lower4pt\hbox{\hskip1pt$\sim$}}
   \raise1pt\hbox{$<$}}}         
\def\gsim{\mathrel{\rlap{\lower4pt\hbox{\hskip1pt$\sim$}}
 \raise1pt\hbox{$>$}}}         

\begin{document}
\draft
\title{Light Quark Energy Distribution in Heavy Quark Symmetry}
\author{Adam Szczepaniak,  Chueng-Ryong Ji and Stephen R. Cotanch}
\address{Department of Physics, North Carolina State University,
Raleigh, North Carolina 27695-8202} \date{\today} \maketitle
\begin{abstract}
Using heavy quark symmetry we construct an energy sum rule relating
the form factor of a heavy flavored meson $\xi(w)$ to the light quark energy
distribution.
We find that the current
 available data for $\xi(w)$ is consistent with a broad energy distribution 
rather than with that of a pure quark/anti-quark bound
state, indicating a large nonvalence content.
\end{abstract}
\narrowtext

There is growing interest in the physics of heavy
quark systems stimulated by the recognition~\cite{hist,ISWI}
of the important role of the heavy quark symmetry of the QCD
lagrangian in the infinite quark mass limit.
Interest has also been fostered by recent measurements
of many heavy hadron decay modes~\cite{exp} which provide
more stringent challenges of our understanding of the interplay
 between strong and weak
interactions~\cite{mix}.
 In the heavy quark mass limit, $m \to
\infty$, the spectrum of heavy mesons containing one heavy and one light
valence quarks, become degenerated and independent
of the polarization state of the light degrees of freedom~\cite{ISWI}.
 Furthermore, an effective decoupling of the dynamics of a heavy quark
  reduces the problem to
that of a cloud of light quarks and  gluons in an external
colored source~\cite{ISWI,SH,RAD}. These two facts imply definite relations
between heavy meson matrix elements of various currents that involve
heavy quark field operators. In particular it has
been shown that in the limit of an exact heavy quark symmetry
all form factors
associated with such matrix elements are determined  by a single,
universal function
$\xi(w)$, called the Isgur-Wise form factor~\cite{ISWI,ISWI1}. The
function $\xi(w)$
is usually introduced through a matrix element  of a vector heavy quark
current evaluated between two $J^P=0^-$ heavy meson states,

\begin{equation}
\langle p'|{\overline \Psi}(0)\gamma^\mu \Psi(0) |p\rangle
= f_+(t)(p'+p)^\mu + f_-(t)(p'-p)^\mu, \label{e1}
\end{equation}
with $p'^2=M'^2,\,p^2=M^2$ and $t=(p'-p)^2$. In the $m,M,M'\to \infty$ limit
it is useful to introduce the
velocity transfer, $t^2/m^2 = (v' - v)^2$.
Defining
$w \equiv v\cdot v' = 1 - t/2m^2$ the form factors $f_\pm(t)$, which are not
independent reduce to a single function of $w$,

\begin{eqnarray}
f_+(t) & \to & { {M'+M}\over \sqrt{4M'M} } \xi(w) \to \xi(w), \nonumber
\\
f_-(t) & = &  { {M-M'}\over {M + M'} }f_(t) \to 0.
\label{e2}
\end{eqnarray}
There have been many attempts to calculate the universal function
$\xi(w)$~\cite{RAD,SCO,BALL} and also the $O(1/m)$
corrections to $f_\pm$~\cite{1overm,pert}.
As already pointed out, due to the decoupling of the heavy quark, $\xi(w)$
is primarily determined by the
dynamics of light degrees of freedom thereby considerably simplifying the
problem. This point is implicit in the
constituent quark model calculations~\cite{SCO} in which $\xi(w)$
is determined by
a noncovariant wave function of a light spectator quark.
Another approach is based on a QCD sum rule analysis
of the vacuum three-point correlator~\cite{RAD}. The standard
interpolation
method is used to relate the spectral representation of the correlator to
the contribution from the ground state, thus to $\xi(w)$.
In such an approach the decoupling of the heavy quark
makes $\xi(w)$ depend on the propagator of the light quark
in an external field generated by the heavy quark.

In this paper we are interested in studying the
structure of the distribution
of light degrees of freedom in the heavy meson, and propose
an  approach in which we relate $\xi(w)$ to the
nondiagonal correlator of two currents, i.e. with only one of the meson
states in Eq.~(\ref{e1}) being interpolated.
 The correlator
is then evaluated between the vacuum and the other heavy meson state. It
is in a sense
an intermediate approach between the constituent quark model and
a three-point function QCD sum rule calculation, and it will become
clear that such an approach
will allow us to relate, in a covariant way, the universal function $\xi(w)$
 to gauge invariant matrix elements describing light
quark and gluon content of a heavy meson.
The general form of a correlator which we shall be dealing with in the rest
of this paper is
\begin{equation}
\delta_{ij}T(p,q) =   i\int dz
e^{iqz} \langle
p,i|TO({z\over 2})O_j(-{z\over 2})|0\rangle,
\label{e3}
\end{equation}
where $|p,i\rangle$ is a heavy meson state with momentum $p$, ($p^2=M^2$)
containing a light valence quark with flavor $i$, and the
operators $O,O_i$ are given by
$O(z) = {\overline \Psi}(z)\Gamma\Psi(z)$,
$O_i(z) = {\overline \Psi}(z)\Gamma' \psi_i(z)$.
Here $\Psi$ and $\psi_i$ are the heavy and light quark field operators
and $\Gamma$ and $\Gamma'$ are some combination of Dirac gamma matrices,
respectively. We start with the phenomenological analysis of the
correlator $T(p,q)$.
The 2-dimensional plane in variables $q^2_{1,2}$, ($q_{1,2} = p/2 \pm q$),
is conventionally parameterized in terms of $\omega,\lambda$ and  $q^2$,
i.e. $q^2_{1,2} = (1\pm\omega)q^2 \pm \lambda + { {p^2}\over 4}$.
For fixed $\omega$ and $\lambda$, with $|\omega| \le 1$ and $\omega
\lambda > p^2/4$,
 $T(p,q) = T(\omega,\lambda,q^2)$ satisfies a dispersion relation in the
variable $q^2$~\cite{disp} with

\begin{equation}
\delta_{ij}Im T(\lambda,\omega,q^2) = {1\over 2} \int dz 
e^{iqz} \left[ \langle
p,i|O({z\over 2})O_j(-{z\over 2})|0\rangle 
+ \langle p,i|O_j(-{z\over 2})O({z\over 2})|0\rangle\right]. \label{e6}
\end{equation}
Inserting a complete set of intermediate states with
masses $M_n$ in Eq.~(\ref{e6}) gives,

\begin{eqnarray}
& & \delta_{ij}T(\lambda,\omega,q^2) \nonumber \\
& & = \sum_n { {\theta \left( {{ M_n^2 - \lambda - p^2/4} \over {1+\omega} }
- s_{min} \right)} \over { M_n^2 - \lambda - p^2/4 - (1+\omega)q^2 -
i\epsilon} } \langle p,i|O(0)|n,q_{1n} \rangle
\langle n,q_{1n} |O_j(0) |0 \rangle_{q^2_{1n}=(p/2+q_n)^2=M^2_n} \nonumber
\\ & & + { {\theta \left( {{ M_n^2 + \lambda - p^2/4} \over
{1-\omega} }
- s_{min} \right)} \over { M_n^2 + \lambda - p^2/4 -
(1-\omega)q^2-i\epsilon} } \langle p,i|O_j(0)|n,q_{2n} \rangle
\langle n,q_{2n} |O(0) |0 \rangle_{q^2_{2n}=(p/2-q_n)^2=M^2_n}. \nonumber \\
\label{e7} \end{eqnarray}
where $s_{min} = \mbox{min} \left( {{-\lambda-p^2/4} \over {1+\omega}},
{{\lambda - p^2/4} \over {1-\omega}} \right)$.
With our choice of the quark content for the operators $O,O_j$
the first term in the RHS of Eq.~(\ref{e7}) has contributions
from resonances containing one heavy and one light valence quark,
 while the second term involves
states
with two heavy quarks and does not contribute to $\xi(w)$. Therefore, we
shall choose the variables
$\omega,\lambda,q^2$ in such a way that the second term is nonleading in the
$1/m$ expansion. This corresponds to the choice $\lambda \gsim p^2/4 $,
($\omega > 0$) and $q^2 \sim p^2/2(1+\omega)$ which leads to a
 suppression by a factor of $O(1/m)$ of the second term with
respect to the first term in Eq.~(\ref{e7}).
Furthermore it can be shown that the
$(1-\omega)$ dependence is also nonleading so that
 in the leading order analysis we set
 $\lambda \gsim p^2/4$ and $\omega = 1$.
It is also more convenient to use, in the infinite heavy quark mass
limit~\cite{sur,cz},
the spectral representation in the energy
variables instead of the invariant mass. With the resonance fixed at
$q^2_{1n} = M^2_n$ the momentum transfer is $t = 2 m^2(1 - w) =(p-q_{1n})^2=
- \lambda
+ p^2/4$ and the correlator will be evaluated for $q^2_1 = (m + E_q)^2
\sim m^2 + 2mE_q$.
Denoting the energy of an excited state by $E_n$, $M_n^2 \sim
m^2 + 2mE_n$, the correlator in the variables $E_q$ and $w$
 for $\omega=1$ and $\lambda \gsim p^2/4$ is given by

\begin{eqnarray}
\delta_{ij}T(w,E_q) =
\sum_n  { {
\langle p,i|O(0)|n,q_{1n} \rangle \langle n,q_{1n} |O_j(0) |0
\rangle_{q_{1n}^2=M_n^2,(p-q_{1n})^2=2m^2(1-w)} }
\over {2m(E_n-E_q-i\epsilon)} }.
 \label{e9}
\end{eqnarray}
In the following we restrict our analysis to two choices of the
operators $O,O_j$ given by two sets of the $\Gamma,\Gamma'$
 matrices;
$\mbox{Set 1} :  \Gamma = \gamma^\mu,\; \Gamma' = \gamma_5$,
$\mbox{Set 2} :  \Gamma = 1,\; \Gamma' = \gamma_5$.
We will explicitly show that to leading order in
the $1/m$ expansion
the two sets lead to a unique sum rule for the form factor $\xi(w)$.
For $|E_q - E_n | >> 0$  a standard assumption is
 that the contributions to the sum in Eq.~(\ref{e9}) from the ground state
and from higher resonances
can be parameterized by a single state with energy $E_0$ and zero
width together with a
 continuum starting at with threshold energy $E_c$
For two $\Gamma$ sets the leading contribution to
the phenomenological correlator
as $m\to \infty$ 
is given by

\begin{eqnarray}
\mbox{Set 1} & : &  T^\mu(w,E_q) = {i\over 2}({3\over 2}p +
q)^\mu
\left[ {f_h\xi(w)\over {(E_q-E_0+i\epsilon)}} + T_c(
\tilde{t},E_q,E_c) \right], \nonumber \\
\mbox{Set 2} & : & T(w,E_q) = {i\over 2} m
 \left[ {f_h(1+w)\xi(w)\over {(E_q-E_0+i\epsilon)}} + T_c(
\tilde{t},E_q,E_c) \right], \label{e10}
\end{eqnarray}
where $f_{h}$ is a ground state heavy meson decay constant,
 $E_c$ is the effective
continuum threshold energy, and $T_c$ is the contribution from the
continuum whose explicit form will be discussed later. In deriving the second
equation above we have used
$\langle p'|{\overline \Psi}(0)\Psi(0) |p\rangle
= m (1 + w)\xi(w)$,
which follows from Eq.~(\ref{e1}) taking the derivative of the
vector current in the $M'\to M$ limit.

We shall now discuss the theoretical description of the correlators.
For $E_q >> E_0 \sim \Lambda_{QCD}$ the heavy quark at the intermediate
state in the correlator in Eq.~(\ref{e3}) is off its energy shell roughly by
$E_q$ and
the singularity of the perturbative heavy quark propagator dominates 
the spacetime
behavior of the operator product in Eq.~(\ref{e3}).
In a fixed gauge, the leading perturbative contribution is given
by~\cite{MUTA}
\begin{equation}
\langle p,i|T{\overline \Psi}({z\over
2})\Gamma\Psi({z\over 2}){\overline \Psi}(-{z\over
2})\Gamma'\psi_j(-{z\over 2}) |0\rangle
\to \delta_{ij}\mbox{Tr}\left[ \langle p,i|{\overline
\Psi}({z\over 2})\psi_i(-{z\over 2})|0\rangle \Gamma S(z) \Gamma'\right],
\label{e11}
\end{equation}
where $S(z)$ is the perturbative heavy quark propagator and the
trace is taken over the spinor indices. The nonperturbative part of the
propagator leads to a $O(1/m)$ correction and is not taken into
account in the leading order analysis. For the two sets of operators
discussed earlier the correlators are given by

\begin{eqnarray}
T^\mu(p,q)  & =  &  if_h\int
d^4z\int {{d^4k}\over {(2\pi)^4}}
{{e^{-i(q-k)z}}\over {k^2 - m^2 + i\epsilon}}
\biggl[ ( k^\mu + p^\mu ) m \phi(z^2,z\cdot p) \nonumber \\
  & + &  k^\mu m \phi_-(z^2,z\cdot p)
     +  k^\nu m \phi_{T\nu\mu}(z^2,z\cdot p)
\biggr], \nonumber \\
T(p,q) &  = & if_h \int d^4z \int
{{d^4k}\over {(2\pi)^4}} {{e^{-i(q-k)z}}
\over {k^2 - m^2 + i\epsilon}}
\left[(k\cdot p + m^2) \phi(z^2,z\cdot p) + m^2
\phi_-(z^2,z\cdot p) \right], \nonumber \\ \label{e12}
\end{eqnarray}
where
\begin{eqnarray}
\phi^{\dag}(z^2,z\cdot p) & \equiv & {-i\over {f_h}}\langle 0|{\overline
\psi}_i({z\over 2}) {{\pbar}\over {p^2}}\gamma_5 \Psi(-{z\over
2})|p,i\rangle, \nonumber \\
\phi_-^{\dag}(z^2,z\cdot p) & \equiv & {i\over {f_h }} \left[
{1\over m} \langle 0|{\overline\psi}_i({z\over 2}) \gamma_5
\Psi(-{z\over 2})|p,i\rangle
+\langle 0|{\overline\psi}_i({z\over 2}) {{\pbar}\over {p^2}}\gamma_5
\Psi(-{z\over 2})|p,i\rangle
\right],
\nonumber \\
\phi_{T\nu\mu}^{\dag}(z^2,z\cdot p) & \equiv & {-i\over {f_h m}}
\langle 0|{\overline\psi}_i({z\over 2}) i\sigma_{\nu\mu}\gamma_5
\Psi(-{z\over 2})|p,i\rangle, \label{e12.1}
\end{eqnarray}
and the normalization of the leading amplitude, $\phi(z^2,z\cdot p)$,
is given by
$\phi(z^2,z\cdot p)_{z = 0} = 1$.
In the infinite heavy quark mass limit the operators that depend on a spin
structure of the light quark (i.e, ${\overline \Psi}\gamma^\mu\gamma_5
\psi_i$ and ${\overline \Psi}\gamma_5 \psi_i$) are
degenerate~\cite{ISWI1}
and  thus lead to the same matrix element. For this reason $\phi_-$
does  not contribute in the leading order.
Similarly it is seen that $\phi_T$ is suppressed by a power of
$1/m$ relative to $\phi$ so only terms proportional to $\phi$
will be retained in the leading order analysis.
The matrix element which defines $\phi$ can be rewritten as

\begin{equation}
\phi^{\dag}(z^2, z\cdot p) =
{1\over { i f_h }} e^{i{p\over 2}\cdot z} \langle 0| \overline{\psi}_i(z)
{{\pbar}\over {p^2}}\gamma_5  \Psi(0) |p,i \rangle
  \equiv  e^{i{p\over 2}\cdot z}\phi^{\dag}_q(z_T^2,z_L) \label{e13}
\end{equation}
with the subscript $L$ indicating the magnitude of a 4-vector projection
parallel to $p^\mu$
 and subscript $T$ indicating the remaining three
components of a 4-vector perpendicular to $p^\mu$.
The structure of the
light quark wave function,
$\phi_q$ is revealed by expanding in a Taylor
series at $z=0$,
\begin{equation}
\phi_q(z_T^2,z_L) = \sum_n { {(iz_L \sqrt{p^2})^n}\over {n!} }
\left[  \langle \phi_q^n \rangle(0)
 + z_T^2  \langle \phi_q^n \rangle'(0) + {{z_T^4}\over 2}
 \langle \phi_q^n \rangle''(0) + \cdots \right]. \label{e15.1}
\end{equation}
The moments $\langle \phi_q^n \rangle^{(m)}(0)$ can be
related to local matrix elements with $n$ longitudinal, $D_L$, and $2m$
transverse, $D_T$, covariant derivative
insertions, $D^\mu = \partial^\mu + ig A^\mu(z)$~\cite{cz}.
If we imagine that the heavy meson is composed of a single light quark
 bound by a chromoelectric potential
of a static heavy quark, i.e. with no dynamical gluons, then $D_L = \partial_L + ig A_L(z_T)$
being $z_L$ independent implies that $\langle \phi_q^n \rangle^{(m)}(0)$
with fixed $m$ determine the energy (longitudinal component of the
4-vector) distribution of a single light quark.
The heavy meson having a definite
energy, $E_0$, requires the light quark to be on the energy shell as well,
and as a consequence $\langle \phi_q^n \rangle(0) = (E_0/\sqrt{p^2})^n$
or

\begin{equation}
\phi_q(z_T^2,z_L) = e^{iz_LE_0}\phi_q(z^2_T).
\end{equation}
The transverse amplitude $\phi_q(z^2_T)$ is the bound state wave function
which
has a nontrivial behavior for $|z_T| \sim 1/E_0$ even in the
static case and it 
approaches a plane wave solution only in a free case i.e. when
$A(z_T) \to 0$. 
However the $z_L$ dependent amplitude is 
the same whenever $\phi_q$ describes a bound state or a free particle.
Since in general the vector potential $A=A(z_L,z_T)$ is
nonstatic one expects a sizable
contributions from the gluon and/or ${q\overline q}$ sea.
The $z_L$ dependence of $A_L$
implies that the spectral representation of $\phi_q(z_L)$
is a smeared distribution around $E=E_0$ rather then being
proportional to a delta function. 
Furthermore since the gluon field $A^\mu = A^\mu(z)$ should be in general
$p^\mu$ independent, 
the division 
between the longitudinal and transverse wave functions is  
purely a  matter of convenience. Also relevant is the scale governing the
energy and momentum smearing intervals
 of the two wave functions which
should be of the same order. Thus, despite the fact that in leading order
 the correlators in Eq.~(\ref{e12}) are not sensitive
 to the transverse wave function, the above argument still permits
 conclusions on the size of the gluon momentum distribution from the
analysis of the longitudinal wave function alone.

Transforming to the momentum representation the correlators are
given by

\begin{eqnarray}
T^\mu(w,E_q)  & =  &  {i\over 2}f_h ({3\over 2}p + q)^\mu
\int_0 {{dE}\over {2\pi}}
{{\phi_q(E)} \over {E_q - E w + i\epsilon}},
\nonumber \\
T(w,E_q) &  = &  {i\over 2} f_h m (1 + w)
\int_0 {{dE}\over {2\pi}}
{{\phi_q(E)}\over  {E_q - E w + i\epsilon}}
 \label{e21}
\end{eqnarray}
with $\phi_q(E) =\int dz_L e^{iE z_L}\phi(z^2_T=0,z_L)$ and
$\int_0 {{dE}\over {2\pi}} \phi_q (E) = 1.$
The leading order theoretical correlators of Eq.~(\ref{e21}) are also used
to define
a continuum contribution $T_c$ introduced in Eq.~(\ref{e10}). The standard
assumption is that
the spectral density of $T_c(\tilde{t},E_q,E_c)$ representing the sum
over higher resonances can be replaced by the theoretical one
 at $E > E_c$.
From Eqs.~(\ref{e10}) and ~(\ref{e21}) we obtain a single sum rule for
$\xi(w)$.

\begin{equation}
\int_0 {{dE}\over {2\pi}} { {\phi_q(E)}\over { E_q - E  w  + i\epsilon} }
 =  { {\xi(w)} \over {E_q - E_0 + i\epsilon} }
+ \int_{ {{E_c}\over w}} {{dE}\over {2\pi}}
 {{\phi_q(E)}\over { E_q - E w  + i\epsilon} }. \label{e24}
\end{equation}
In order to reduce the contributions from the unknown higher order
terms
in the expansion in Eq.~(\ref{e15.1}) and suppress the sensitivity of the sum
rule to the phenomenological parameterization of the spectral density, a
standard Borel transformation ($E_q \to T$) is performed for
Eq.~(\ref{e24}) leading to

\begin{eqnarray}
& & e^{-w E_0/T} +
\int_0 {{dE}\over {2\pi}} e^{-w E/T} \left[ \phi_q(E) - 2\pi\delta(E-E_0)
\right] \nonumber \\
& & =  \xi(w)e^{-E_0/T}
+ \int_{ {{E_c}\over w}}
 {{dE}\over {2\pi}} e^{-w E/T} \phi_q(E ). \label{e25}
\end{eqnarray}
As explained earlier in a simple quantum mechanical description in which a
heavy meson contains a single quark orbiting around a static
chromoelectric source,
 $\phi_q(E) = 2\pi\delta(E - E_0)$ and since $E_c > E_0$,
the sum rule automatically leads to a correct normalization, $\xi(1)=1$, for
all values of $T$. At first look the above result may seem dependent on the
choice of the parameterization of the phenomenological spectral density.
This is however not the case. If a more detailed parameterization were used,
involving explicitly higher resonances instead of a smooth
continuum
 approximation, due to orthogonality of the mass eigenstates such
resonances would
not contribute to the RHS of Eq.~(\ref{e25}) at $w=1$ leaving $\xi(1) = 1$
coming only from the ground state contribution.
In a realistic situation $\phi_q(E)$ is a smeared distribution
around $E=E_0$. Furthermore it follows from the analysis 
of the normalization, $f_{H\to L}$, of the heavy-to-light 
meson matrix elements that the $f_{H\to L} \sim O(m^{-3/2})$ 
behavior ~\cite{cz1} requires $\phi_q(E)$ to vanish only linearly with  
$E$ for $E/E_0 << 1$~\cite{asnew}. In order to be consistent with 
the phenomenological parameterization of the spectral density 
we shall assume that $\phi_q(E)$ vanishes for $E > E_c$.
The value of the threshold energy has been obtained in Ref.~\cite{RAD,cz} 
 and it varies in the range $2E_0 \lsim  E_c \lsim 3E_0$.
We then use the following parameterization 
$\phi_q \propto E(E_c - E)\exp(-(E-E_0)^2/E_W^2)$ and study the 
form factor $\xi(w)$ for different values of the  width parameter 
$E_W$. For a given value of $E_c/E_0$ and $E_W/E_0$ the
Borel parameter ($T/E_0$) is fixed by the normalization 
$\xi(1) = 1$. 
In Fig.~1 we plot our predictions for $\xi(w)$ for 
$E_c/E_0 = 2.1,2.5,3$ given by the set of dashed, solid and 
dotted curves respectively and for $E_W/E_0 = .1\mbox{ and }10$
 corresponding to upper and lower curves in each set
(dashed, solid, dotted) respectively.
As $E_W$ decreases the predictions are sensitive to both the detailed
shape of the energy distribution of the light quark and to the
value
of the continuum threshold energy. On the other hand, for increasing 
$E_W$ our predictions become weakly dependent on $E_c/E_0$ and
$E_W/E_0$, and are bound from below by $1/w^2$ behavior which in turn
follows from the linear behavior of $\phi_q(E)$ at small $E$.
The data corresponding to the $B \to D$ form factor (finite $m$) and
extracted
from the $B\to Dl{\overline \nu}_l$ with $A=(\tau_B/1.18)*(|V_{cb}|/0.05)
= 1.11$ have been taken from Ref.~\cite{data}.
The comparison of our results with the experimental data 
shows that the $\xi(w)$ form factor is inconsistent with the usually
assumed peaking ($\delta$-type)
approximation to the distribution amplitude corresponding to small
 $E_W/E_0$)~\cite{example}
and suggests
a rather broad, $E_W > E_0$, energy distribution. This in turn implies
large gluon and possibly sea quark amplitudes in a heavy
meson and we
conclude that a large fraction of the energy-momentum
of the light degrees of freedom is distributed over the nonvalence
components.
Although our results rely on a comparison with heavy, but finite, quark mass
data, both the
 $O(1/m)$
and perturbative, $O(\alpha(E))$ corrections, as shown in
Ref.~\cite{pert}, are expected to be very small not exceeding
a few percent. 

\acknowledgments
A.S. would like to thank Stephen Sharpe and the INT staff for the
hospitality during
 the {\it Phenomenology and
Lattice}
summer program where part of this work was done.
Financial support from U.S. D.O.E. grants
DE-FG05-88ER40461 and DE-FG05-90ER40589 is also acknowledged.

\begin{figure}
\caption{The Isgur-Wise form factor $\xi(w)$. The dash-doted line is the
$1/w^2$ asymptotic behavior. The variuos theoretical curves are explainded
in the text.}

\end{figure}

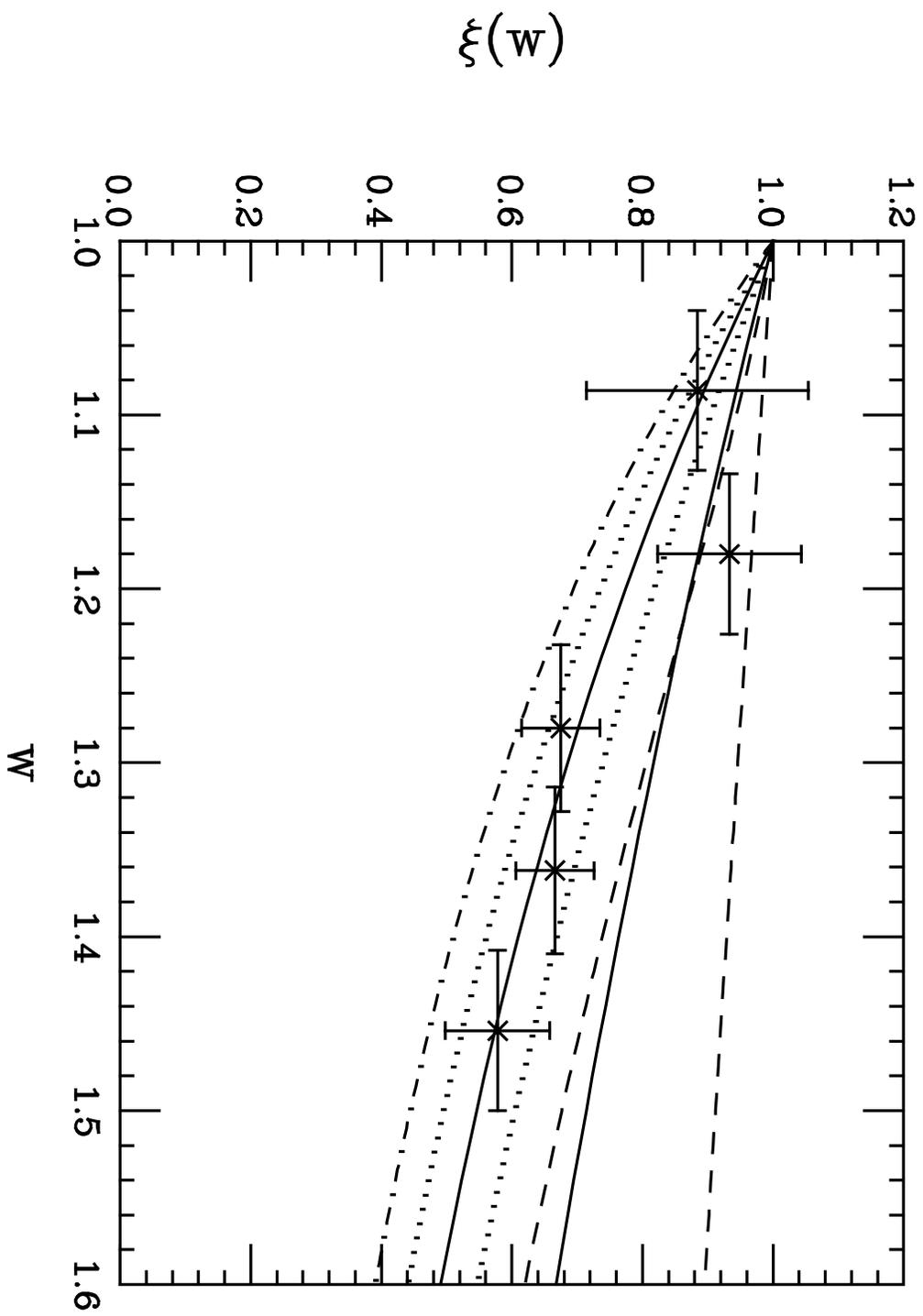
\end{document}